\documentclass[aps]{revtex4}

\usepackage{amsmath}
\usepackage{amssymb}
\usepackage{stmaryrd}
\usepackage{natbib}
\numberwithin{equation}{section}
\usepackage{graphicx}

\usepackage{epsfig}

\def\ov#1{\overline{#1}}

\def\vb#1{\mbox{\boldmath$#1$}}
\def\pd#1#2{\frac{\partial #1}{\partial #2}}

\def\bdot{\,\vb{\cdot}\,}

\def\cal#1{\mathcal{#1}}

\newcommand{\bc}{\begin{center}}
\newcommand{\ec}{\end{center}}
\newcommand{\bt}{\begin{tabbing}}
\newcommand{\et}{\end{tabbing}} 
\newcommand{\be}{\begin{eqnarray*}}
\newcommand{\ee}{\end{eqnarray*}}
\newcommand{\bs}{\begin{slide}}
\newcommand{\es}{\end{slide}}

\begin{document}

\title{Lie-transform derivation of oscillation-center quasilinear theory}

\author{Alain J.~Brizard$^{1,a}$}
\affiliation{$^{1}$Department of Physics, Saint Michael's College, Colchester, Vermont 05439, USA \\
$^{a}$ {\rm Author to whom correspondence should be addressed: abrizard@smcvt.edu}}

\date{\today}
\def\corrEmail{abrizard@smcvt.edu}

\begin{abstract}
The derivation of the oscillation-center quasilinear theory in an unmagnetized plasma by Dewar \cite{Dewar:1973} is rederived by Lie-transform perturbation method. New results not included in Dewar's original paper are rigorously derived and the conservation laws of energy and momentum, which combine the contributions of the resonant and non-resonant particles, are presented in both particle phase space and oscillation-center phase space.
\end{abstract}

\maketitle


\section{Introduction}

The problem of the quasilinear diffusion of a background Vlasov distribution in the presence of electromagnetic waves is a classic example of wave-particle interactions beyond the linear regime \cite{Davidson:1972,Stix:1992,Kaufman_Cohen:2019}. Because the nonlinear Vlasov-Maxwell equations can be represented in terms of a Hamiltonian system, in which energy and momentum are conserved exactly at all orders in wave amplitude, it is important to formulate quasilinear theory in terms of resonant particles, which allow the transfer of energy and momentum to the waves, and non-resonant particles which ensure that total (waves+particles) energy and momentum are conserved. In a classic paper, Kaufman \cite{Kaufman:1972} iteratively solved the Vlasov-Poisson equations for an unmagnetized plasma  up to second order, to obtain a resonant first-order Vlasov contribution and a second-order nonresonant Vlasov contribution which yielded exact energy-momentum conservation laws.

In a ground-breaking paper published in 1973, Dewar \cite{Dewar:1973} rederived Kaufman's quasilinear theory \cite{Kaufman:1972} by constructing a Hamiltonian theory of quasilinear theory in an unmagnetized plasma, in which resonant and non-resonant particles contribute to the conservation of energy and momentum in the presence of quasilinear diffusion of resonant particles associated with electrostatic fluctuations. This derivation was based on the canonical transformation from particle phase space $({\bf x},{\bf p})$  to oscillation-center phase space $({\bf X},{\bf P})$, which is generated by the  generating function $S({\bf x},{\bf P},t)$  \cite{Goldstein:2001}: ${\bf X} = {\bf x} + \partial S({\bf x},{\bf P},t)/\partial{\bf P}$ and ${\bf p} = {\bf P} + \partial S({\bf x},{\bf P},t)/\partial{\bf x}$, where the Hamilton-Jacobi perturbation equation
\begin{equation}
\pd{}{t}S({\bf x},{\bf P},t) \;+\; h\left({\bf x}, {\bf P} + \pd{S}{\bf x},t\right) \;=\; H\left({\bf x} + \pd{S}{\bf P},{\bf P},t\right),
\label{eq:HJ_eq}
\end{equation}
is solved iteratively in powers of the perturbation wave amplitude $\delta$, where $h$ and $H$ denote the particle Hamiltonian and the oscillation-center Hamiltonian, respectively. 

In the absence of quasilinear diffusion, the oscillation-center phase-space transformation \cite{Dodin_Fisch:2010} removes the fast (quiver) dynamics of a charged particle moving in a high-frequency electromagnetic field with a weakly-nonuniform amplitude. The resulting slow oscillation-center (quasiparticle) dynamics involves motion in the presence of a second-order ponderomotive potential $H_{2}({\bf X},{\bf P},t)$ whose phase-space derivatives yield the oscillation-center Hamilton equation $d{\bf X}/dt = {\bf P}/m + \delta^{2}\partial H_{2}/\partial{\bf P}$ and $d{\bf P}/dt = -\,\delta^{2}\partial H_{2}/\partial{\bf P} \equiv {\bf F}_{\rm pond}$. Since the high-frequency ponderomotive potential is normally inversely proportional to the particle's mass [see Eq.~\eqref{eq:pond_Ham}], we immediately find that the electron species is predominantly affected by the ponderomotive force ${\bf F}_{\rm pond}$. In the theory of high-frequency laser-plasma interactions \cite{Bobin:1985}, for example, the ponderomotive force plays an important role in the process of laser-plasma acceleration \cite{Krushelnick:2010,Kaw:2017,Tajima:2020}. In low-frequency gyrokinetic theory \cite{Brizard_Hahm:2007}, on the other hand, ponderomotive effects are predominantly affected by ions. In both cases, the ponderomotive potential yield polarization and magnetization effects that play crucial role in their respective reduced plasma dynamics \cite{Cary_Kaufman:1977,Cary_Kaufman:1981,Brizard:2009}.

In the presence of quasilinear diffusion, Dewar \cite{Dewar:1973} solves the Hamilton-Jacobi perturbation equation \eqref{eq:HJ_eq} up to first order in wave amplitude, and shows that the energy-momentum conservation laws derived by Kaufman \cite{Kaufman:1972} are still valid. The interested reader may wish to read the paper by Krommes \cite{Krommes:2012} on the historical impact of Dewar's 1973 paper. In addition, the work of Kaufman in plasma wave theory is reviewed by Tracy and Brizard \cite{Tracy_Brizard:2009}, while the mathematical foundations of quasilinear theory in inhomogeneous plasma are presented by Dodin \cite{Dodin:2022,Dodin:2024} in two recent review papers. In another recent paper, Brizard and Chan \cite{Brizard_Chan:2022} constructed the Hamiltonian formulation of quasilinear diffusion in a uniform magnetized plasma that is perturbed by electromagnetic waves, and demonstrated that their results were identical to the quasilinear theory of Kennel and Engelmann \cite{K_E}.

The purpose of the present paper is to rederive Dewar's oscillation-center quasilinear theory for an unmagnetized plasma in the presence of an electrostatic wave where the canonical transformation from particle phase space to oscillation-center phase space is carried by Lie-transform perturbation method \cite{Kaufman:1978,Cary_Kaufman:1981,Brizard:2009}. This perturbation method is superior to perturbatively solving the Hamilton-Jacobi perturbation equation \eqref{eq:HJ_eq} since it can be systematically carried out to arbitrary order in $\delta$, as was demonstrated from its applications in guiding-center theory \cite{Littlejohn:1983,Cary_Brizard:2009,Tronko_Brizard:2015} and gyrokinetic theory \cite{Brizard_Hahm:2007}.

The oscillation-center transformation removes the non-resonant part of the electrostatic wave-particle interactions by constructing the new oscillation-center phase-space coordinates, while the oscillation-center Vlasov equation retains the resonant part of the electrostatic wave-particle interactions. Because of the historical importance of Dewar's work \cite{Dewar:1973}, the paper presents a complete and systematic derivation of the Dewar's oscillation-center quasilinear theory \cite{Dewar:1973}  and, along the way, we also rederive the important energy-momentum conservation laws presented by Kaufman \cite{Kaufman:1972}, where the roles of the resonant and non-resonant Vlasov contributions are explicitly presented. We note that the presentation has a tutorial style, which will enable the reader to acquire a deeper understanding of Dewar's oscillation-center quasilinear theory \cite{Dewar:1973}. In addition, while the Hamilton-Jacobi perturbation equation \eqref{eq:HJ_eq} offers no guidance on how to proceed beyond the first order in perturbation amplitude, the use of Lie-transform perturbation theory provides a systematic pathway to go beyond quasilinear theory.

\section{Oscillation-center phase-space transformation}

The oscillation-center phase-space transformation \cite{Brizard:2009} from particle phase space $z^{\alpha} = ({\bf x},{\bf p},w,t)$ to oscillation-center phase space $Z^{\alpha} = ({\bf X},{\bf P},W,t)$ is expressed in terms of the canonical perturbation expansion
\begin{equation}
    Z^{\alpha} \;\equiv\; {\sf T}_{\delta}z^{\alpha} \;=\; \cdots T_{2}\,T_{1}z^{\alpha} \;=\; z^{\alpha} \;+\; \delta\,\left\{ S_{1},\; z^{\alpha}\right\} \;+\; \delta^{2}\,\left\{ S_{2},\; z^{\alpha}\right\} \;+\; 
    \frac{\delta^{2}}{2} \left\{ S_{1} ,\frac{}{} \{ S_{1},\; z^{\alpha}\}\right\} + \cdots,
    \label{eq:Z_z}
\end{equation}
where $\delta$ denotes the order of the perturbation, with $T_{n} \equiv \exp(\delta^{n}\{S_{n},\;\})$ defined in terms of the generating scalar field $S_{n}$, which generates the canonical phase-space transformation up to an arbitrary order in $\delta$. Here, the canonical Poisson bracket $\{\;,\;\}$ is defined in eight-dimensional extended phase space as
\begin{equation}
    \{ f,\; g\} \;\equiv\; \left(\pd{f}{w}\,\pd{g}{t} \;-\; \pd{f}{t}\,\pd{g}{w}\right) \;+\; \left(\nabla f\bdot\pd{g}{\bf p} \;-\; 
    \pd{f}{\bf p}\bdot\nabla g \right),
    \label{eq:PB}
\end{equation}
where the six-dimensional canonical coordinates $({\bf x},{\bf p})$ have been extended to include the canonical energy-time coordinates $(w,t)$. Up to second order in the perturbation amplitude, the oscillation-center coordinates are expressed as
\begin{eqnarray}
    {\bf X} &=& {\bf x} \;-\; \delta\,\pd{S_{1}}{\bf p} \;-\; \delta^{2}\,\pd{S_{2}}{\bf p} \;-\; \frac{\delta^{2}}{2}\;\left\{ S_{1},\; \pd{S_{1}}{\bf p} \right\}, \\
    {\bf P} &=& {\bf p} \;+\; \delta\,\nabla S_{1} \;+\; \delta^{2}\,\nabla S_{2} \;+\; \frac{\delta^{2}}{2}\;\left\{ S_{1},\frac{}{} \nabla
    S_{1} \right\}, \\
    W &=& w \;-\; \delta\,\pd{S_{1}}{t} \;-\; \delta^{2}\,\pd{S_{2}}{t} \;-\; \frac{\delta^{2}}{2}\;\left\{ S_{1},\; 
    \pd{S_{1}}{t} \right\},
\end{eqnarray}
while time $t$ is not transformed since we are assuming that the generating scalar fields $(S_{1},S_{2},\cdots)$ do not depend on the energy coordinate $w$. We note that, since the perturbation parameter $\delta \ll 1$ is assumed to be small, the phase-space transformation (\ref{eq:Z_z}) is a near-identity transformation, with its inverse easily expressed as
\begin{equation}
    z^{\alpha} \;=\; Z^{\alpha} \;-\; \delta\,\left\{ S_{1},\; Z^{\alpha}\right\} \;-\; \delta^{2}\,\left\{ S_{2},\; Z^{\alpha}\right\} \;+\; 
    \frac{\delta^{2}}{2} \left\{ S_{1},\frac{}{} \{ S_{1},\; Z^{\alpha}\} \right\} + \cdots,
    \label{eq:z_Z}
\end{equation}
where the canonical Poisson bracket (\ref{eq:PB}) that appears on the right side is now expressed in terms of the oscillation-center coordinates $({\bf X},{\bf P},W,t)$. We also note that the sign convention for the Lie-transform perturbation method adopted here \cite{Brizard:2009}, which differs from Dewar's sign convention [see Dewar's Eqs.~(2)-(3), henceforth referred to as Eqs.~\cite{Dewar:1973}(2)-(3)], is consistent with the perturbation method used to derive the gyrokinetic Vlasov-Maxwell equations \cite{Brizard_Hahm:2007}.

\subsection{Oscillation-center Hamiltonian perturbation equation}

The purpose of the oscillation-center transformation (\ref{eq:Z_z}) is to transform the extended particle Hamiltonian
\begin{equation}
    h({\bf x},{\bf p},w,t) \;=\; \frac{|{\bf p}|^{2}}{2m} \;+\; e\,\delta\,\Phi_{1}({\bf x},t) \;-\; w,
    \label{eq:Ham_p}
\end{equation}
where $\Phi_{1}({\bf x},t)$ denotes the perturbation electrostatic potential (labeled as first order in the perturbation amplitude), into the extended oscillation-center Hamiltonian
\begin{equation}
    H({\bf X},{\bf P},W,t) \;=\; H_{0}({\bf P},W) \;+\; \delta\,H_{1}({\bf X},{\bf P},t) \;+\; \delta^{2}\,H_{2}({\bf X},{\bf P},t) \;+\; \cdots,
    \label{eq:Ham_oc}
\end{equation}
where $H_{1}({\bf X},{\bf P},t) \equiv 0$ in the absence of wave-particle resonances (e.g., when a uniform plasma is perturbed by a high-frequency wave \cite{Cary_Kaufman:1977,Cary_Kaufman:1981}). The transformation from Eq.~(\ref{eq:Ham_p}) to Eq.~(\ref{eq:Ham_oc}) is induced by the phase-space transformation (\ref{eq:Z_z}) through the {\it push-forward} operator ${\sf T}_{\delta}^{-1}$, which yields the following Hamiltonian perturbation equation
\begin{eqnarray}
    H &\equiv& {\sf T}_{\delta}^{-1}h \;=\; h \;-\; \delta\,\{ S_{1},\; h\} \;-\; \delta^{2}\,\left\{ S_{2},\; h\right\} \;+\; \frac{\delta^{2}}{2} \left\{ S_{1},\frac{}{} \{ S_{1},\; h\} \right\} + \cdots \nonumber \\
    &=& H_{0} \;+\; \delta\,H_{1} \;+\; \delta^{2}\,H_{2} \;+\; \cdots,
    \label{eq:Ham_pert}
\end{eqnarray}
where
\begin{eqnarray}
    H_{0} &=& h_{0}({\bf P},W) \;=\; \frac{|{\bf P}|^{2}}{2m} \;-\; W, 
    \\
    H_{1} &=& e\,\Phi_{1}({\bf X},t) \;-\; \{S_{1},\; h_{0}\} \;\equiv\; e\,\Phi_{1}({\bf X},t) \;-\; \frac{d_{0}S_{1}}{dt}, 
    \label{eq:OCH_1} \\
    H_{2} &=& -\,e\;\{ S_{1},\; \Phi_{1}\} \;-\; \{ S_{2},\; h_{0}\} 
    \;+\; \frac{1}{2} \left\{ S_{1},\frac{}{} \{ S_{1},\; h_{0}\} \right\} \;\equiv\; -\,e\;\{ S_{1},\; \Phi_{1}\} \;-\; \frac{d_{0}S_{2}}{dt} \;+\; \frac{1}{2} \left\{ S_{1},\; \frac{d_{0}S_{1}}{dt} \right\},
    \label{eq:OCH_2} 
\end{eqnarray}
where we have introduced the unperturbed total time derivative 
\[ \frac{d_{0}}{dt} \equiv \{ \;,\; h_{0}\} = -\;\pd{h_{0}}{w}\;\pd{}{t} \;+\; \pd{h_{0}}{\bf p}\bdot\nabla = \pd{}{t} + {\bf v}\bdot\nabla, \]
expressed in terms of the particle velocity ${\bf v} \equiv {\bf p}/m$. 

Lastly, we wish to connect the Hamiltonian perturbation equation \eqref{eq:Ham_pert} to the nonperturbative Hamilton-Jacobi equation previously derived by Johnston and Kaufman \cite{Johnston_Kaufman:1978}. First, by evaluating the perturbation derivative of Eq.~\eqref{eq:Ham_pert}, we obtain 
\begin{equation}
\frac{dH}{d\delta} \;=\; \left(\frac{d}{d\delta} {\sf T}_{\delta}^{-1}\right)h \;+\; {\sf T}_{\delta}^{-1}\left(\frac{dh}{d\delta}\right) \;\equiv\; \left[\left(\frac{d}{d\delta}{\sf T}_{\delta}^{-1}\right){\sf T}_{\delta}\right]\,H \;+\; {\sf T}_{\delta}^{-1}\left(\frac{dh}{d\delta}\right), 
\label{eq:Delta_Ham}
\end{equation}
where 
\begin{equation}
\left[\left(\frac{d}{d\delta}{\sf T}_{\delta}^{-1}\right){\sf T}_{\delta}\right] \;=\; -\,\left( {\cal L}_{1} \;+\; 2\delta\,{\cal L}_{2} \;+\; 3\,\delta^{2}{\cal L}_{3} \;+\; \cdots \right) \;+\; \delta^{2} \left({\cal L}_{2}\,{\cal L}_{1} \;-\; {\cal L}_{1}\,{\cal L}_{2}\right) \;+\; \cdots.
\label{eq:d_Tdelta}
\end{equation}
Second, we apply the operator \eqref{eq:d_Tdelta} on the transformed Hamiltonian $H$ to obtain
\[ \left[\left(\frac{d}{d\delta}{\sf T}_{\delta}^{-1}\right){\sf T}_{\delta}\right]H \;=\; -\; \left\{ \left( S_{1} \;+\; 2\delta\,S_{2} \;+\; 3\,\delta^{2}S_{3} \;+\; \cdots \right),\; H \right\} \;+\; \delta^{2} \left( \left\{ S_{2},\;\{S_{1},\; H\}\right\} \;-\; \left\{ S_{1},\;\{S_{2},\; H\}\right\}\right) \;+\; \cdots, \]
so that, if we use the Jacobi identity 
\[ \left\{ S_{2},\;\{S_{1},\; H\}\right\} \;-\; \left\{ S_{1},\;\{S_{2},\; H\}\right\} \;=\; \left\{ S_{2},\;\{S_{1},\; H\}\right\} \;+\; \left\{ S_{1},\;\{H,\; S_{2}\}\right\} \;=\; -\;\left\{ H,\; \{S_{2},\; S_{1}\}\right\} \;=\; \left\{ \{S_{2},\; S_{1}\},\; H \right\}, \]
Eq.~\eqref{eq:Delta_Ham} becomes the perturbation Hamilton-Jacobi equation in oscillation-center phase space $({\bf X},{\bf P},W,t)$:
\begin{equation}
-\;\left[\left(\frac{d}{d\delta}{\sf T}_{\delta}^{-1}\right){\sf T}_{\delta}\right]H \;\equiv\; {\cal L}_{\cal S}H \;=\; \{ {\cal S},\; H\} \;=\; {\sf T}_{\delta}^{-1}\left(\frac{dh}{d\delta}\right)  \;-\; \frac{dH}{d\delta}
\label{eq:HJ_pert}
\end{equation}
where ${\cal S} \equiv S_{1} + 2\delta\,S_{2} + \delta^{2} \left( 3\,S_{3} - \{ S_{2},\; S_{1}\} \right) + \cdots$ defines the Lie generating function \cite{Johnston_Kaufman:1978}, which nonperturbatively generates the canonical  transformation from particle phase-space coordinates to oscillation-center phase-space coordinates. This perturbation Hamilton-Jacobi equation is analogous to the Hamilton-Jacobi equation appearing in Dewar's work \cite{Dewar:1973}, which is expressed in mixed phase-space coordinates $({\bf x},{\bf P})$. We note that, when Eq.~\eqref{eq:HJ_pert}  is expanded up to the first two orders in the perturbation amplitude, we obtain the first-order perturbation equation \eqref{eq:OCH_1} and  the second-order perturbation equation \eqref{eq:OCH_2}, respectively.

\subsection{Oscillation-center Vlasov equation}

Once the oscillation-center phase-space transformation has been constructed to a desired order (e.g., $\delta^{2}$), we may now transform the perturbed particle Vlasov equation
\begin{equation}
    \frac{d_{0}f}{dt} \;=\; e\,\delta\,\{\Phi_{1},\; f\},
    \label{eq:Vlasov_p}
\end{equation}
where the particle Vlasov distribution $f({\bf x},{\bf p},t)$ is now transformed to the oscillation-center Vlasov distribution $F({\bf X},{\bf P},t)$ through the {\it pull-back} operator ${\sf T}_{\delta}$:
\begin{equation}
    f \;\equiv\; {\sf T}_{\delta}F \;=\; F \;+\; \delta\,\{ S_{1},\; F\} \;+\; \delta^{2}\,\left\{ S_{2},\; F\right\} \;+\; \frac{\delta^{2}}{2} \left\{ S_{1},\frac{}{} \{ S_{1},\; F\} \right\} + \cdots \;=\; f_{0} \;+\; \delta\,f_{1} \;+\; \delta^{2}\,f_{2} \;+\; \cdots,
\end{equation}
where the oscillation-center Vlasov distribution is expanded as $F = F_{0} + \delta\,F_{1} + \delta^{2}\,F_{2} + \cdots$, so that
\begin{eqnarray}
    f_{0} &=& F_{0}, \label{eq:fF_0} \\
    f_{1} &=& F_{1} \;+\; \{ S_{1},\; F_{0}\}, \label{eq:f1} \\
    f_{2} &=& F_{2} \;+\; \{ S_{1},\; F_{1}\} \;+\; \{ S_{2},\; F_{0}\} \;+\; \frac{1}{2} \left\{ S_{1},\frac{}{} \{ S_{1},\; F_{0}\} \right\}. \label{eq:fF_2} 
\end{eqnarray}
Since the oscillation-center Vlasov equation is often expected to be easier to solve (e.g., without wave-particle resonances), then its solution $F = F_{0} + \delta\,F_{1} + \cdots$ can be used to generate the solution $f = f_{0} + \delta\,f_{1} + \cdots$ of the particle Vlasov equation.

By combining the pull-back and push-forward operations, the oscillation-center transformation of the particle Vlasov equation (\ref{eq:Vlasov_p}) yields the oscillation-center Vlasov equation
\begin{equation}
    {\sf T}_{\delta}^{-1}\left(\frac{d_{0}}{dt}{\sf T}_{\delta}F \right) \;\equiv\; \frac{d_{\delta}F}{dt} \;=\; e\,\delta\,\left\{ {\sf T}_{\delta}^{-1}\Phi_{1},\; F \right\} \;=\; e\,\delta\,
    \left\{ \left(\Phi_{1} \;-\; \delta\,\{S_{1},\; \Phi_{1}\} + \cdots \right), \frac{}{} F \right\},
\end{equation}
where the canonical oscillation-center Poisson bracket $\{\;,\;\}$ is obtained from Eq.~(\ref{eq:PB}) with a simple change of coordinates (i.e., all canonical Poisson brackets have the same structure), and the oscillation-center time-evolution operator $d_{\delta}/dt$ is defined, up to second order in $\delta$, as
\begin{eqnarray}
   \frac{d_{\delta}F}{dt} &=& \frac{d_{0}}{dt} \left( F \;+\; \delta\,\{ S_{1},\; F\} \;+\; \delta^{2}\,\left\{ S_{2},\; F\right\} \;+\; \frac{\delta^{2}}{2} \left\{ S_{1},\frac{}{} \{ S_{1},\; F\} \right\} \right) \nonumber \\
    &&-\; \delta\,\left\{ S_{1},\frac{}{} \frac{d_{0}}{dt}\left( F \;+\frac{}{} \delta\,\{ S_{1},\; F\} \right) \right\} - \delta^{2} 
    \left\{ S_{2},\; \frac{d_{0}F}{dt} \right\} + \frac{\delta^{2}}{2} \left\{ S_{1},\; \left\{ S_{1},\; \frac{d_{0}F}{dt} \right\} \right\}\nonumber \\
   &=& \frac{d_{0}F}{dt} \;+\; \left\{ \left(
   \delta\,\frac{d_{0}S_{1}}{dt} + \delta^{2}\,\frac{d_{0}S_{2}}{dt}
   - \frac{\delta^{2}}{2} \left\{ S_{1},\; \frac{d_{0}S_{1}}{dt}\right\}
   \right),\; F \right\}.
   \label{eq:d_delta}
\end{eqnarray}
In deriving this final expression for Eq.~(\ref{eq:d_delta}), we have used the Jacobi identity. Since the canonical Poisson bracket has constant coefficients, we also used the identity $(d_{0}/dt)\{f,\; g\} \equiv \{ d_{0}f/dt,\; g\} + \{ f,\; d_{0}g/dt\}$.

Finally, we can now write the oscillation-center Vlasov equation at each perturbation order, from zeroth order $(\delta^{0})$ up to second order $(\delta^{2})$:
\begin{eqnarray}
    \frac{d_{0}F_{0}}{dt} &=& 0, \label{eq:OCVlasov_0} \\
    \frac{d_{0}F_{1}}{dt} &=& \left\{ \left(e\,\Phi_{1} \;-\; \frac{d_{0}S_{1}}{dt}\right),\; F_{0}\right\} \;\equiv\; \{ H_{1},\; F_{0} \}, \label{eq:OCVlasov_1} \\ 
    \frac{d_{0}F_{2}}{dt} &=& \left\{ \left(e\,\Phi_{1} - \frac{d_{0}S_{1}}{dt}\right), F_{1}\right\} + \left\{ \left( \frac{1}{2} \left\{ S_{1}, \frac{d_{0}S_{1}}{dt}\right\} - \frac{d_{0}S_{2}}{dt} - e\,\{ S_{1}, \Phi_{1}\} \right), F_{0} \right\} \equiv \{ H_{1}, F_{1}\} + \{ H_{2}, F_{0}\},
     \label{eq:OCVlasov_2}
\end{eqnarray}
where the first-order and second-order oscillation-center Hamiltonians $(H_{1},H_{2})$ are defined in Eqs.~(\ref{eq:OCH_1})-(\ref{eq:OCH_2}), respectively. Hence, the time evolution of the oscillation-center Vlasov distribution $F$ only depends on the oscillation-center Hamiltonian $H$.

\subsection{Oscillation-center Vlasov theory in the absence of quasilinear diffusion}

In the absence of quasilinear diffusion in a uniform unmagnetized plasma (i.e., without wave-particle resonances), the unperturbed Vlasov background is space- and time-independent,  and Fourier analysis can be carried out explicitly. First, the perturbation electrostatic potential can be expanded as $\Phi_{1}({\bf x},t) \equiv \ov{\Phi}_{1}\,\exp[i\,\theta({\bf x},t)] + \ov{\Phi}_{1}^{*}\,
\exp[-i\,\theta({\bf x},t)]$, where $\theta({\bf x},t) \equiv {\bf k}\bdot{\bf x} - \omega\,t$, while the first-order scalar field can be expanded as $S_{1}({\bf x},{\bf p},t) \equiv \ov{S}_{1}({\bf p})\,\exp[i\,\theta({\bf x},t)] + \ov{S}_{1}^{*}({\bf p})\,
\exp[-i\,\theta({\bf x},t)]$. Hence, the first-order perturbation equation \eqref{eq:OCH_1} can be written as 
\begin{equation}
\frac{d_{0}S_{1}}{dt} \;=\; i\,\frac{d_{0}\theta}{dt} \left( e^{i\theta}\,\ov{S}_{1}({\bf p}) \;-\; e^{-i\theta}\,\ov{S}^{*}_{1}({\bf p}) \right) \;=\; e\,  \left( e^{i\theta}\,\ov{\Phi}_{1} \;+\; e^{-i\theta}\,\ov{\Phi}^{*}_{1} \right) \;-\; H_{1},
\label{eq:OCH_1_H1}
\end{equation}
where $d_{0}\theta/dt = -\,(\omega - {\bf k}\bdot{\bf v})$. Since the oscillation-center transformation is formally introduced to remove the wave phase $\theta$ from the reduced oscillation-center dynamics, it is clear that the first-order oscillation-center Hamiltonian $H_{1} \equiv 0$ vanishes identically since the remaining terms in Eq.~\eqref{eq:OCH_1_H1} are all explicitly dependent on the wave phase. The solution for the first-order generating function is, therefore, expressed as \cite{Cary_Kaufman:1977,Cary_Kaufman:1981,Brizard:2009} 
\begin{equation}
    \ov{S}_{1}({\bf p}) \;=\; \frac{i\,e\,\ov{\Phi}_{1}}{(\omega - {\bf k}\bdot{\bf v})}.
    \label{eq:S1_0}
\end{equation}
In addition, since $H_{1} = 0$, the first-order oscillation-center Vlasov equation \eqref{eq:OCVlasov_1} becomes $d_{0}F_{1}/dt = 0$.

At second order in the perturbation analysis, the second-order perturbation can be written as
\begin{equation}
\frac{d_{0}S_{2}}{dt} \;=\; 2i\,\frac{d_{0}\theta}{dt} \left( e^{2i\theta}\,\ov{S}_{2}({\bf p}) \;-\; e^{-2i\theta}\,\ov{S}^{*}_{2}({\bf p}) \right) \;=\; -\,\frac{e}{2} \left\{ S_{1},\; \Phi_{1}\right\} \;-\; H_{2},
\label{eq:OCH_2_H2}
\end{equation}
where the purpose of the second-order generating function $S_{2}({\bf x},{\bf p},t) \equiv \ov{S}_{2}({\bf p})\,\exp[2i\,\theta({\bf x},t)] + \ov{S}_{2}^{*}({\bf p})\,\exp[-2i\,\theta({\bf x},t)]$ is to remove the second wave-phase harmonic from the oscillation-center dynamics. Because the second-order term $-\,(e/2) \left\{ S_{1},\; \Phi_{1}\right\}$ contains a term that is independent of the wave phase $\theta$, the second-order oscillation-center Hamiltonian is defined in terms of its wave-phase average \cite{Cary_Kaufman:1977,Cary_Kaufman:1981,Brizard:2009}
\begin{equation}
    H_{2}({\bf P}) \;\equiv\; -\,\left.\left.\frac{1}{2}\right\langle \left\{ S_{1},\;e\,\Phi_{1}\right\}\right\rangle \;=\; \frac{i}{2}\,e{\bf k}\bdot \left( \ov{\Phi}_{1}\,\pd{\ov{S}_{1}^{*}}{\bf P} \;-\; \ov{\Phi}_{1}^{*}\,
    \pd{\ov{S}_{1}}{\bf P} \right) \;=\; 
    \frac{e^{2}|{\bf k}|^{2}\,|\ov{\Phi}_{1}|^{2}}{m\,(\omega - {\bf k}\bdot{\bf v})^{2}},
    \label{eq:pond_Ham}
\end{equation}
where $\langle\cdots\rangle$ denotes an average with respect to the wave phase $\theta$, while the second-order perturbation equation \eqref{eq:OCH_2_H2} becomes
\begin{equation}
\frac{d_{0}S_{2}}{dt} \;=\; -\,\left.\left.\left.\left.\frac{e}{2}\right(  \left\{ S_{1},\; \Phi_{1}\right\} \;-\; \right\langle \left\{ S_{1},\; \Phi_{1}\right\}\right\rangle\right),
\label{eq:OCH_2_S2}
\end{equation}
whose solution yields the second-harmonic wave amplitude 
\begin{equation}
\ov{S}_{2}({\bf p}) \;=\; -\;\;\frac{e^{2}|{\bf k}|^{2} \ov{\Phi}_{1}^{2}}{2m\,(\omega - {\bf k}\bdot{\bf v})^{2}}.
\end{equation}
The second-order oscillation-center Vlasov equation \eqref{eq:OCVlasov_2} becomes $d_{0}F_{2}/dt = \{H_{2},\;F_{0}\}$, so that the second-order oscillation-center Vlasov distribution $F_{2}$ evolves under the influence of the second-order ponderomotive Hamiltonian.

In particle phase space, the first-order particle Vlasov distribution \eqref{eq:f1} is
\begin{equation}
    f_{1} \;=\; \nabla S_{1}\bdot\pd{F_{0}}{\bf p} \;=\; \left( 
    \ov{S}_{1}\,e^{i\theta} \;-\; \ov{S}_{1}^{*}\,e^{-i\theta}\right) i\,{\bf k}\bdot\pd{F_{0}}{\bf p} \;=\; -e\,\left( \ov{\Phi}_{1}\,e^{i\theta} \;+\; \ov{\Phi}_{1}^{*}\,e^{-i\theta}\right) 
    \frac{{\bf k}}{(\omega - {\bf k}\bdot{\bf v})}\bdot
    \pd{F_{0}}{\bf p},
\end{equation}
where $F_{1} = 0$ in the absence of quasilinear diffusion. When it is inserted into the perturbed Poisson equation
\begin{equation}
    |{\bf k}|^{2}\ov{\Phi}_{1} \;=\; 4\pi\,e \int \ov{f}_{1}\,d^{3}p \;=\; -\;4\pi\,e^{2}\,\ov{\Phi}_{1}\,\int \frac{{\bf k}}{(\omega - {\bf k} \bdot{\bf v})}\bdot\pd{F_{0}}{\bf p} \;d^{3}p,
    \label{eq:Poisson}
\end{equation}
where summation over particle species is implied, we obtain the dispersion relation expressed in terms of the permittivity function
\begin{equation}
    \epsilon({\bf k},\omega) \;\equiv 1 \;+\; \frac{4\pi\,e^{2}}{|{\bf k}|^{2}}\,\int \frac{{\bf k}}{(\omega - {\bf k} \bdot{\bf v})}
    \bdot\pd{F_{0}}{\bf p} \;d^{3}p \;=\; 0.
    \label{eq:epsilon}
\end{equation}
In what follows, we denote $\omega_{0}({\bf k})$ as one of the roots of the dispersion relation (\ref{eq:epsilon}), and we will assume a monochromatic electrostatic wave with $({\bf k},\omega_{0}({\bf k}))$ for simplicity of presentation. Otherwise, the amplitudes $\ov{\Phi}_{1}({\bf k},\omega_{i}({\bf k}))$ are functions of ${\bf k}$ on a specific root $\omega_{i}({\bf k})$ of the dispersion relation, and final results in particle and oscillation-center phase spaces are obtained by integrating over wave-number space ($\int d^{3}k/(2\pi)^{3}$) and summing 
$(\sum_{i})$ over the roots of the dispersion relation. We forgo including these technical details into our presentation in order to preserve the simplicity of the presentation; the Reader is invited to refer to Kaufman's \cite{Kaufman:1972} and Dewar's \cite{Dewar:1973} for these details.

\section{Oscillation-center Vlasov theory in the presence of quasilinear diffusion} 

In the presence of quasilinear diffusion, the background Vlasov distribution $f_{0}({\bf p},\tau)$ acquires a slow time dependence that scales with the second order in the perturbation parameter $\delta$ (i.e., $\tau \equiv \delta^{2}t$). Hence, $d_{0}f_{0}/dt \equiv \delta^{2}\,\partial f_{0}/\partial\tau$ and the slow time-dependence of $f_{0}$ implies that the root $\omega_{0}({\bf k},\tau)$ of the dispersion relation (\ref{eq:epsilon}) also has a slow time dependence. Hence, while the background unmagnetized plasma is still spatially uniform, it is no longer time independent and Fourier analysis can only be carried in space.

\subsection{Time-dependent eikonal Theory}

The linear time-dependent Vlasov equation can be solved iteratively by using a multiple-time scale method developed by Bateman and Kruskal \cite{Bateman_Kruskal:1972}. The asymptotic solution proposed by Bateman and Kruskal is presented as follows. First, we consider the partial differential equation
\begin{equation}
    \pd{f(u,\delta\,t)}{t} \;-\; i\,D(\omega,u)\,f(u,\delta\,t) \;=\; g(u,\delta\,t),
    \label{eq:fg_ODE}
\end{equation}
where both functions $(f,g)$ depend on the coordinate $u$ and are weakly dependent on time $(\delta \ll 1)$, while $D \equiv \omega - u$ is assumed to be time independent. By introducing the integrating factor $\exp(-\,i\,D\,t)$, we obtain the formal solution
\begin{equation}
    e^{-\,i\,D\,t} \;f(u,\delta\,t) \;=\; \int_{-\infty}^{t} e^{-i\,D\,t'}\;g(u,\tau^{\prime})\,dt',
\end{equation}
where the ``initial'' value of the first-order field is $f(u,-\infty) \equiv 0$, and we assume that $\omega$ has a small positive imaginary part so that $\exp(-i\,D\,t) \rightarrow 0$ as $t \rightarrow -\,\infty$. After performing two integration by parts, this formal solution becomes
\begin{equation}
    e^{-\,i\,D\,t} \;f(u,\delta\,t) = e^{-\,i\,D\,t} \left( \frac{i\,g}{D} + \frac{1}{D^{2}}\;\pd{g}{t} \right) - \frac{1}{D^{2}} \int_{-\infty}^{t} 
    e^{-i\,D\,t'}\;\frac{\partial^{2}g(u,\delta\,t')}{\partial t'^{2}}\,dt' = e^{-\,i\,D\,t} \left( g + i\,\pd{g}{t}\,\pd{}{\omega}\right)\frac{i}{D} \;+\; \cdots,
\end{equation}
where we, henceforth, ignore second-order partial time derivatives on $g(u,\delta\,t)$, and we have expressed $D^{-2} \equiv -\,\partial D^{-1}/\partial\omega$. Hence, the solution of Eq.~(\ref{eq:fg_ODE}) is expressed as
\begin{equation}
    f(u,\delta\,t) \;=\; \left( g(u,\delta\,t) \;+\; i\,\pd{g(u,\delta\,t)}{t}\,\pd{}{\omega}\right)\frac{i}{\omega - u},
    \label{eq:fg_sol}
\end{equation}
up to first order in partial time derivative. This general solution, which includes time-dependence up to first order in the perturbation parameter $\delta \ll 1$, will be used prominently in what follows when solving the first-order oscillations-center equations for $\ov{S}_{1}({\bf p},\delta\,t)$ and $\ov{F}_{1}({\bf p},\delta\,t)$.

\subsection{Linear oscillation-center Vlasov equation}

We now return the first-order oscillation-center perturbation equation \eqref{eq:OCH_1}. Since the time dependence of the background Vlasov distribution is weak, we can proceed with a time-dependent eikonal representation of the perturbation potential 
\begin{equation} 
    \Phi_{1}({\bf x},t) \;\equiv\; \ov{\Phi}_{1}(\delta t)\,
    \exp[i\,\Theta({\bf x},t)] + \ov{\Phi}_{1}^{*}(\delta t)\,
    \exp[-i\,\Theta({\bf x},t)],
\end{equation}
where the eikonal phase is defined as $\Theta({\bf x},t) = {\bf k}\bdot{\bf x} - \int_{t_{0}}^{t}\omega_{0}({\bf k},\delta^{2}t')\,dt'$, while the first-order scalar field is represented as
\begin{equation} 
    S_{1}({\bf x},{\bf p},t) \;\equiv\; \ov{S}_{1}({\bf p},\delta t)\,
    \exp[i\,\Theta({\bf x},t)] + \ov{S}_{1}^{*}({\bf p},\delta t)\,
    \exp[-i\,\Theta({\bf x},t)].
\end{equation}
In this case, the first-order oscillation-center Hamiltonian $\ov{H}_{1} \neq 0$ no longer vanishes. Instead, the resonant first-order oscillation-center Hamiltonian is defined as
\begin{equation}
    \ov{H}_{1} \;\equiv\; e\,\ov{\Phi}_{1}\,\Delta \;+\; i\,e\,\pd{\ov{\Phi}_{1}}{t}\;\pd{\Delta}{\omega_{0}} \;=\; e\,\ov{\Phi}_{1} \;+\; i\,(\omega_{0} - {\bf k}\bdot{\bf v})\,\ov{S}_{1} 
    \;-\; \pd{\ov{S}_{1}}{t} ,
    \label{eq:H1_S1}
\end{equation}
where the window function  \cite{Dewar:1973}
\begin{equation}
\Delta(\omega_{0} - {\bf k}\bdot{\bf v}) \;\equiv\; {\sf H}\left( \frac{1}{2}\,\Delta\omega \;-\; |\omega_{0} - {\bf k}\bdot{\bf v}|\right) \;=\; \left\{ \begin{array}{lr}
1 & (|\omega_{0} - {\bf k}\bdot{\bf v}| < \frac{1}{2}\,\Delta\omega) \\
 & \\
 0 &  (|\omega_{0} - {\bf k}\bdot{\bf v}| > \frac{1}{2}\,\Delta\omega)
 \end{array}\right.
\end{equation}
is defined in terms of the Heaviside step function ${\sf H}$, with the window width $\Delta\omega$ assumed to be much smaller than any spectral features of the perturbed electrostatic potential $\Phi_{1}$. Hence, Eq.~(\ref{eq:H1_S1}) yields the differential equation for the non-resonant first-order oscillation-center scalar field $\ov{S}_{1}$:
\begin{equation}
    \pd{\ov{S}_{1}}{t} \;-\; i\,(\omega_{0} - {\bf k}\bdot{\bf v})\,\ov{S}_{1} \;=\; e\,\ov{\Phi}_{1}\,(1 - \Delta) \;+\; i\,e\,\pd{\ov{\Phi}_{1}}{t}\;
    \pd{(1 - \Delta)}{\omega_{0}}.
\end{equation}
By using the formal solution (\ref{eq:fg_sol}), we readily find the solution
\begin{eqnarray}
    \ov{S}_{1} &=& \left[e\,\ov{\Phi}_{1}\,(1 - \Delta) \;+\; i\,e\,\pd{\ov{\Phi}_{1}}{t}\;\pd{(1 - \Delta)}{\omega_{0}}\right] 
    \frac{i}{\omega_{0} - {\bf k}\bdot{\bf v}} \;-\; e\,\pd{\ov{\Phi}_{1}}{t}\,(1 - \Delta)\,\pd{}{\omega_{0}}\left(\frac{1}{\omega_{0} - {\bf k}\bdot{\bf v}}\right) 
    \nonumber \\
    &\equiv& \left( e\,\ov{\Phi}_{1} \;+\; i\,e\,\pd{\ov{\Phi}_{1}}{t}\,\pd{}{\omega_{0}}\right) \frac{i\,(1 - \Delta)}{(\omega_{0} - {\bf k}\bdot{\bf v})},
    \label{eq:S1_Delta}
\end{eqnarray}
where we omit the second-order partial time derivative of $\ov{\Phi}_{1}$. We note that Eq.~\eqref{eq:S1_Delta} is identical to Eq.~\cite{Dewar:1973}(24) except for the sign difference mentioned above. As we approach the resonance at $\omega_{0} = {\bf k}\bdot{\bf v}$, the numerator $1 - \Delta$ vanishes and, therefore, the scalar field $\ov{S}_{1}({\bf p},\delta t)$ vanishes at resonance.

Next, when Eq.~(\ref{eq:H1_S1}) is inserted into the first-order oscillation-center Vlasov equation (\ref{eq:OCVlasov_1}), we obtain
\begin{equation}
    \pd{\ov{F}_{1}}{t} \;-\; i\,(\omega_{0} - {\bf k}\bdot{\bf v})\,\ov{F}_{1} \;=\; i\,{\bf k}\bdot\pd{F_{0}}{\bf p} \left[ e\,\ov{\Phi}_{1}\,\Delta \;+\; i\,e\,\pd{\ov{\Phi}_{1}}{t}\;\pd{\Delta}{\omega_{0}}\right],
\end{equation}
which yields the resonant first-order oscillation-center Vlasov distribution
\begin{equation}
    \ov{F}_{1} \;=\; -\,{\bf k}\bdot\pd{F_{0}}{\bf p} \left( e\,\ov{\Phi}_{1} \;+\; i\,e\,\pd{\ov{\Phi}_{1}}{t}\,\pd{}{\omega_{0}}\right)
    \frac{\Delta}{(\omega_{0} - {\bf k}\bdot{\bf v})},
    \label{eq:F1_OC}
\end{equation}
which is not derived by Dewar \cite{Dewar:1973}. Once again, the second-order partial time derivative of $\ov{\Phi}_{1}$ is omitted. 

\subsection{Poisson equation in the presence of quasilinear diffusion}

We now express the Poisson equation (\ref{eq:Poisson}) in particle phase-space as
\begin{equation}
    |{\bf k}|^{2}\,\ov{\Phi}_{1}(\delta\,t) \;=\; 4\pi\,e \int \ov{f}_{1}({\bf p},\delta\,t)\; d^{3}p
\end{equation}
where the first-order Vlasov distribution is defined as
\begin{equation}
    \ov{f}_{1} \;\equiv\; \ov{F}_{1} \;+\; i\,{\bf k}\bdot\pd{F_{0}}{\bf p}\;\ov{S}_{1} \;=\; -\,{\bf k}\bdot\pd{f_{0}}{\bf p} \left( e\,\ov{\Phi}_{1} \;+\; i\,e\,\pd{\ov{\Phi}_{1}}{t}\,\pd{}{\omega_{0}}\right)\frac{1}{(\omega_{0} - {\bf k}\bdot{\bf v})},
    \label{eq:f1_p}
\end{equation}
where the resonance window function $\Delta$ has canceled out, and once again the second-order particle time derivative is omitted. We note that the solution (\ref{eq:f1_p}), also given by Eq.~\cite{Dewar:1973}(26), exactly matches the solution of the first-order Vlasov equation (\ref{eq:Vlasov_p}) given by Eq.~(\ref{eq:fg_sol}).

When Eq.~(\ref{eq:f1_p}) is inserted into the Poisson equation (\ref{eq:Poisson}), we obtain the dispersion relation
\begin{equation}
    \epsilon({\bf k},\omega_{0})\,\ov{\Phi}_{1} \;+\; i\,
    \pd{\epsilon_{r}({\bf k},\omega_{0})}{\omega_{0}}\;\pd{\ov{\Phi}_{1}}{t} 
    \;=\; 0,
    \label{eq:Poisson_OC}
\end{equation}
where the permittivity $\epsilon({\bf k},\omega_{0})$ is defined in Eq.~(\ref{eq:epsilon}). We now note that the permittivity $\epsilon({\bf k},\omega_{0}) \equiv \epsilon_{r}({\bf k},\omega_{0}) + i\,
\epsilon_{i}({\bf k},\omega_{0})$ can be separated into real and imaginary parts, where $|\epsilon_{i}| \ll \epsilon_{r}$. When this decomposition is inserted into the Poisson equation (\ref{eq:Poisson_OC}), we obtain the real-valued dispersion relation $\epsilon_{r}({\bf k},\omega_{0}) = 0$, and the imaginary-valued relation
\[ \epsilon_{i}({\bf k},\omega_{0})\,\ov{\Phi}_{1} \;+\; 
    \pd{\epsilon_{r}({\bf k},\omega_{0})}{\omega_{0}}\;\pd{\ov{\Phi}_{1}}{t} 
    \;=\; 0, \]
which yields the field evolution equation
\begin{equation}
    \pd{\ov{\Phi}_{1}}{t} \;=\; -\,\left(\pd{\epsilon_{r}}{\omega_{0}}
    \right)^{-1}\epsilon_{i}\,\ov{\Phi}_{1} \;\equiv\; \gamma\,
    \ov{\Phi}_{1},
    \label{eq:Phi1_dot}
\end{equation}
where $\gamma$ denotes the exponential growth rate of the eikonal perturbation amplitude $\ov{\Phi}_{1}$. In particular, by substituting the Plemelj formula \cite{Stix:1992}
\begin{equation}
    \lim_{\nu \rightarrow 0^{+}}\left( \frac{1}{\omega_{0} - {\bf k}
    \bdot{\bf v} + i\,\nu}\right) \;\equiv\; {\cal P}\left( 
    \frac{1}{\omega_{0} - {\bf k}\bdot{\bf v}}\right) \;-\; i\,\pi\;
    \delta(\omega_{0} - {\bf k}\bdot{\bf v})
    \label{eq:Plemelj}
\end{equation}
where ${\cal P}$ denotes the principal part, into the permittivity function (\ref{eq:epsilon}), we find
\[ \epsilon({\bf k},\omega_{0}) \;=\; 1 \;+\; \frac{4\pi\,e^{2}}{|{\bf k}|^{2}} \int {\bf k}\bdot\pd{f_{0}}{\bf p} \left[ {\cal P}\left( 
    \frac{1}{\omega_{0} - {\bf k}\bdot{\bf v}}\right) \;-\; i\,\pi\;
    \delta(\omega_{0} - {\bf k}\bdot{\bf v}) \right] d^{3}p, \]
which immediately yields the (real) dispersion relation 
 \begin{equation}
    \epsilon_{r}({\bf k},\omega_{0}) = 1 \;+\; \frac{4\pi\,e^{2}}{|{\bf k}|^{2}}\,\int {\bf k}\bdot\pd{f_{0}}{\bf p} \left[ {\cal P}\left(
    \frac{1}{(\omega_{0} - {\bf k}\bdot{\bf v})} \right)\right] \;d^{3}p \;=\; 0.
    \label{eq:epsilon_r}
\end{equation}   
while the imaginary part is
\begin{equation}
    \epsilon_{i}({\bf k},\omega_{0}) \;=\; -\;\frac{4\pi\,e^{2}}{|{\bf k}|^{2}} \int {\bf k}\bdot\pd{f_{0}}{\bf p} \left[ \pi\frac{}{}
    \delta(\omega_{0} - {\bf k}\bdot{\bf v})\right] d^{3}p,
\end{equation}
so that the growth rate defined in Eq.~(\ref{eq:Phi1_dot}) becomes
\begin{equation}
    \gamma \;=\; -\,\left(\pd{\epsilon_{r}}{\omega_{0}}
    \right)^{-1}\epsilon_{i} \;=\; \frac{4\pi\,e^{2}}{|{\bf k}|^{2}}\left(\pd{\epsilon_{r}}{\omega_{0}}\right)^{-1} \int {\bf k}\bdot
    \pd{f_{0}}{\bf p} \left[ \pi\frac{}{}\delta(\omega_{0} - {\bf k}\bdot
    {\bf v})\right] d^{3}p,
\end{equation}
which is also given by Eq.~\cite{Dewar:1973}(32). Hence, the wave field eikonal amplitude $\ov{\Phi}_{1}$ grows exponentially as a result of particle energy being transferred from the particles that are resonant with the wave. Once again, we note that the amplitudes 
$\ov{\Phi}_{1}({\bf k},\omega_{i}({\bf k}))$ are functions of ${\bf k}$ on a specific root $\omega_{i}({\bf k})$ of the dispersion relation, and final results in particle and oscillation-center phase spaces are obtained by integrating over wave-number space ($\int d^{3}k/(2\pi)^{3}$) and summing $(\sum_{i})$ over the roots of the dispersion relation.

\section{Quasilinear diffusion in particle phase space}

Before continuing with quasilinear theory in oscillation-center phase space, we present quasilinear theory in particle phase space, and rederive the results presented by Kaufman \cite{Kaufman:1972}.

\subsection{Second-order particle Vlasov equation}

First, we proceed with the derivation of the quasilinear diffusion equation for background Vlasov distribution $f_{0}({\bf p},\tau)$ as well as the derivation of the spatially-averaged second-order particle Vlasov distribution $\langle f_{2}\rangle({\bf p},\delta t)$. These two distributions appear in the second-order particle Vlasov equation
\begin{equation}
    \pd{f_{0}}{\tau} \;+\; \pd{\langle f_{2}\rangle}{t} \;=\; e\left\langle \left\{ \Phi_{1},\frac{}{} f_{1}\right\}\right\rangle \;=\; \pd{}{\bf p}\bdot\left\langle e\,\nabla\Phi_{1}\frac{}{} f_{1}
    \right\rangle \;\equiv\; \pd{}{\bf p}\bdot\left[ e\,i\,{\bf k}\left( \ov{\Phi}_{1}\;\ov{f}_{1}^{*} \;-\frac{}{} \ov{\Phi}_{1}^{*}
    \;\ov{f}_{1}\right) \right],
    \label{eq:f0_2_start}
\end{equation}
where the eikonal amplitude of the first-order particle Vlasov distribution is given by Eq.~(\ref{eq:f1_p}). Upon substituting Eq.~(\ref{eq:f1_p}) into Eq.~(\ref{eq:f0_2_start}), we obtain
\begin{eqnarray}
   \pd{f_{0}}{\tau} + \pd{\langle f_{2}\rangle}{t} &=& \pd{}{\bf p} \bdot \left[ -\,2\,e^{2}|\ov{\Phi}_{1}|^{2}\;{\bf k}{\bf k}\bdot
   \pd{f_{0}}{\bf p} \;{\rm Im}\left( \frac{1}{\omega_{0} - {\bf k}\bdot{\bf v}}\right) \right] \nonumber \\
   &&-\; \pd{}{\bf p} \bdot \left[ e^{2}\,{\bf k}{\bf k}\bdot\pd{f_{0}}{\bf p}\; \pd{|\ov{\Phi}_{1}|^{2}}{t}\; \pd{}{\omega_{0}}{\rm Re}\left( \frac{1}{\omega_{0} - {\bf k}\bdot{\bf v}}\right) \right].
\end{eqnarray}
By using the Plemelj formula (\ref{eq:Plemelj}), we can separate the time evolution of $f_{0}$ and $\langle f_{2}\rangle$ in terms of the quasilinear diffusion equation for the background Vlasov distribution $f_{0}$:
\begin{equation}
    \pd{f_{0}}{\tau} \;=\; \pd{}{\bf p}\bdot\left(\mathbb{D}\bdot
    \pd{f_{0}}{\bf p}\right),
    \label{eq:QL_eq}
\end{equation}
expressed in terms of the quasilinear diffusion tensor \cite{Kaufman:1972}
\begin{equation}
    \mathbb{D} \;\equiv\; e^{2}\,|\ov{\Phi}_{1}|^{2}\;{\bf k}{\bf k} \left[ 2\pi\frac{}{} \delta(\omega_{0} - {\bf k}\bdot{\bf v})\right],
    \label{eq:D_QL}
\end{equation}
which takes into account only those particles that are resonant with the wave, while the space-averaged second-order particle Vlasov distribution is now expressed as
\begin{equation}
    \langle f_{2}\rangle \;=\; -\; \pd{}{\bf p} \bdot \left[ e^{2}|\ov{\Phi}_{1}|^{2}\,{\bf k}{\bf k}\bdot\pd{f_{0}}{\bf p}\; \pd{}{\omega_{0}}{\cal P} \left( \frac{1}{\omega_{0} - 
    {\bf k}\bdot{\bf v}}\right) \right],
    \label{eq:f2_NR}
\end{equation}
which takes into account the non-resonant part of the particle Vlasov distribution. This second-order non-resonant particle Vlasov distribution was derived iteratively by Kaufman \cite{Kaufman:1972} and is given in Eq.~\cite{Dewar:1973}(42), without derivation, by Dewar.  We note that similar expressions have also been derived in App.~B of a recent paper \cite{Dodin:2024} by Dodin.

\subsection{Energy-momentum conservation laws}

The derivation of the conservation laws of energy and momentum proceeds by properly taking into account the separation between the energy-momentum of the waves and the energy-momentum of the particles. 

\subsubsection{Energy conservation law}

The particle energy is simply expressed in terms of the unperturbed kinetic energy of the particles
\begin{equation}
    {\cal E}_{0}(t) \;=\; \int f_{0}({\bf p},\tau)\;\frac{|{\bf p}|^{2}}{2m}\; d^{3}p.
\end{equation}
By using the quasilinear diffusion equation (\ref{eq:QL_eq}), we obtain the time derivative
\begin{eqnarray}
    \frac{d{\cal E}_{0}}{dt} &=& \delta^{2}\,\int \pd{f_{0}}{\tau}({\bf p},\tau)\;\frac{|{\bf p}|^{2}}{2m}\; d^{3}p \;=\; -\,\delta^{2} \int {\bf v}\bdot\mathbb{D}\bdot\pd{f_{0}}{\bf p}\;d^{3}p \nonumber \\
    &=& -\,\delta^{2}\,e^{2}\,|\ov{\Phi}_{1}|^{2}\;2\pi\,\omega_{0}\;
    \int {\bf k}\bdot\pd{f_{0}}{\bf p}\; \delta(\omega_{0} - {\bf k}\bdot{\bf v})\;d^{3}p,
    \label{eq:dE_0}
\end{eqnarray}
where the term ${\bf k}\bdot{\bf v}$ was replaced with $\omega_{0}$ through the resonance condition $\omega_{0} - {\bf k}\bdot{\bf v} = 0$ imposed by the delta function $\delta(\omega_{0} - {\bf k}\bdot{\bf v})$.

Next, the total wave energy, which is expressed as the sum of the wave energy and the contribution from the non-resonant second-order Vlasov distribution:
\begin{eqnarray}
    {\cal E}_{2} &\equiv& \delta^{2} \left[ \frac{|{\bf k}|^{2}}{4\pi}\;
    |\ov{\Phi}_{1}|^{2} \;+\; \int \frac{|{\bf p}|^{2}}{2m}\;\langle f_{2}\rangle\; d^{3}p \right] \nonumber \\
    &=& \delta^{2}\;\frac{|{\bf k}|^{2}}{4\pi}\;|\ov{\Phi}_{1}|^{2} 
    \left[ 1 \;+\; \pd{}{\omega_{0}}\left( \frac{4\pi\,e^{2}}{|{\bf k}|^{2}} \int {\bf k}\bdot\pd{f_{0}}{\bf p}\; \omega_{0}\,{\cal P} \left( \frac{1}{\omega_{0} - {\bf k}\bdot{\bf v}}\right) \right) \right] \nonumber \\&\equiv& \delta^{2}\;\frac{|{\bf k}|^{2}}{4\pi}\;|\ov{\Phi}_{1}|^{2} 
    \left[ 1 \;+\; \pd{}{\omega_{0}}\left(\omega_{0}\frac{}{}(\epsilon_{r} - 1)
    \right) \right] \;=\; \delta^{2}\;\frac{|{\bf k}|^{2}}{4\pi}\;|\ov{\Phi}_{1}|^{2}\;\omega_{0}\;\pd{\epsilon_{r}}{\omega_{0}},
    \label{eq:E2_def}
\end{eqnarray}
where we have used the dispersion relation $\epsilon_{r}({\bf k},\omega_{0}) = 0$. The time derivative of the wave energy is now expressed as
\begin{equation}
    \frac{d{\cal E}_{2}}{dt} \;=\; \delta^{2}\;\frac{|{\bf k}|^{2}}{4\pi}\;\left(2\,\gamma\frac{}{}|\ov{\Phi}_{1}|^{2}\right)\;\omega_{0}\;\pd{\epsilon_{r}}{\omega_{0}} \;=\; \delta^{2}\;|{\bf k}|^{2}\;|\ov{\Phi}_{1}|^{2}\;2\pi\,\omega_{0}\;\int {\bf k}\bdot\pd{f_{0}}{\bf p}\;
    \delta(\omega_{0} - {\bf k}\bdot{\bf v})\;d^{3}p,
\end{equation}
which, when combined with Eq.~(\ref{eq:dE_0}), yields the energy conservation law $d({\cal E}_{0} + {\cal E}_{2})/dt = 0$.

\subsubsection{Momentum conservation law}

The particle momentum is simply expressed in terms of the unperturbed momentum of the particles
\begin{equation}
    \vb{\cal P}_{0}(t) \;=\; \int {\bf p}\;f_{0}({\bf p},\tau)\; d^{3}p.
\end{equation}
By using the quasilinear diffusion equation (\ref{eq:QL_eq}), we obtain the time derivative
\begin{eqnarray}
    \frac{d\vb{\cal P}_{0}}{dt} &=& \delta^{2}\,\int {\bf p}\;\pd{f_{0}}{\tau}({\bf p},\tau)\; d^{3}p \;=\; -\,\delta^{2} \int \mathbb{D}\bdot\pd{f_{0}}{\bf p}\;d^{3}p \nonumber \\
    &=& -\,\delta^{2}\,e^{2}\,|\ov{\Phi}_{1}|^{2}\;2\pi\,{\bf k}\;
    \int {\bf k}\bdot\pd{f_{0}}{\bf p}\; \delta(\omega_{0} - {\bf k}\bdot
    {\bf v})\;d^{3}p
    \label{eq:dP_0}
\end{eqnarray}

Next, the wave momentum is expressed solely as the momentum carried by the non-resonant second-order Vlasov distribution:
\begin{equation}
    \vb{\cal P}_{2} \;\equiv\; \delta^{2} \int {\bf p}\;\langle f_{2}\rangle\; d^{3}p \;=\; \delta^{2}\;\frac{|{\bf k}|^{2}}{4\pi}\;|\ov{\Phi}_{1}|^{2}\;{\bf k}\;\pd{\epsilon_{r}}{\omega_{0}} \;\equiv\;
    \frac{\bf k}{\omega_{0}}\;{\cal E}_{2},
    \label{eq:P2_def}
\end{equation}
which has a simple relation with the total wave energy (\ref{eq:E2_def}). The time derivative of the total wave momentum is now expressed as
\begin{equation}
    \frac{d\vb{\cal P}_{2}}{dt} \;=\; \delta^{2}\;\frac{|{\bf k}|^{2}}{4\pi}\;\left(2\,\gamma\frac{}{}|\ov{\Phi}_{1}|^{2}\right)\;{\bf k}\;\pd{\epsilon_{r}}{\omega_{0}} \;=\; \delta^{2}\;|{\bf k}|^{2}\;|\ov{\Phi}_{1}|^{2}\;2\pi\,{\bf k}\;\int {\bf k}\bdot
    \pd{f_{0}}{\bf p}\;\delta(\omega_{0} - {\bf k}\bdot{\bf v})\;d^{3}p,
    \label{eq:dP_2}
\end{equation}
which, when combined with Eq.~(\ref{eq:dP_0}), yields the momentum conservation law $d(\vb{\cal P}_{0} + \vb{\cal P}_{2})/dt = 0$.

\section{Quasilinear diffusion in Oscillation-center Phase Space} 

We now return to our presentation of quasilinear theory in oscillation-center phase space, which is omitted in Dewar's paper \cite{Dewar:1973}.

\subsection{Second-order oscillation-center Vlasov equation}

The second-order Vlasov equation in oscillation-center phase space is expressed as 
\begin{equation}
    \pd{F_{0}}{\tau} + \frac{d_{0}F_{2}}{dt} \;=\; \{ H_{1},\; F_{1}\} \;+\; \{ H_{2},\; F_{0}\},
\end{equation}
which is unfortunately not discussed by Dewar \cite{Dewar:1973}. By spatially averaging this equation (equivalent to an eikonal average over $\Theta$), we obtain the averaged second-order oscillation-center Vlasov equation
\begin{equation}
    \pd{F_{0}}{\tau} + \pd{\langle F_{2}\rangle}{t} \;=\; \left\langle\{ H_{1},\frac{}{} F_{1}\}\right\rangle \;+\; \{ \langle H_{2}\rangle,\; F_{0}\} \;=\; \pd{}{\bf P}\bdot\left\langle \nabla H_{1}\frac{}{} F_{1} \right\rangle \;=\; \pd{}{\bf P}\bdot\left[ i\,{\bf k}\,\left(\ov{H}_{1}\,\ov{F}_{1}^{*} \;-\frac{}{} \ov{H}_{1}^{*}\,\ov{F}_{1}\right) \right],
\end{equation}
where $\langle H_{2}\rangle \equiv K_{2}({\bf P})$ is the ponderomotive Hamiltonian (\ref{eq:pond_Ham}) and thus $\{ \langle H_{2}\rangle, F_{0}\}
\equiv 0$. First, we note that the surviving terms $(\ov{H}_{1},\ov{F}_{1})$ on the right side are only associated with resonant wave-particle contributions. 

In the following analysis, we will use the Plemelj formula (\ref{eq:Plemelj}) to obtain the resonant and non-resonant identities
\begin{equation}
    \left. \begin{array}{rcl}
    \Delta/(\omega_{0} - {\bf k}\bdot{\bf v}) &\equiv& -i\,\pi\;\delta(\omega_{0} - {\bf k}\bdot{\bf v}) \\
     && \\
    (1 - \Delta)/(\omega_{0} - {\bf k}\bdot{\bf v}) &\equiv& {\cal P}\left[(\omega_{0} - {\bf k}\bdot{\bf v})^{-1}\right]
    \end{array} \right\}.
\end{equation}
Up to first order in partial-time derivatives, we find
\begin{eqnarray} 
\ov{H}_{1}\,\ov{F}_{1}^{*} &=& \left[ e\,\ov{\Phi}_{1}\,\Delta \;+\; i\,e\,\pd{\ov{\Phi}_{1}}{t}\;\pd{\Delta}{\omega_{0}} \right] \left[ -\,{\bf k}\bdot\pd{F_{0}}{\bf P} \left( e\,
\ov{\Phi}_{1}^{*} \;-\; i\,e\,\pd{\ov{\Phi}_{1}^{*}}{t}\,\pd{}{\omega_{0}}\right)\frac{\Delta}{(\omega_{0} - {\bf k}\bdot{\bf v})^{*}}\right] \nonumber \\
 &=& -\,{\bf k}\bdot\pd{F_{0}}{\bf P} \left[ \left( e^{2}\,|\ov{\Phi}_{1}|^{2} \;+\; i\,e^{2}\,\ov{\Phi}_{1}^{*}\,\pd{\ov{\Phi}_{1}}{t}\,\pd{\Delta}{\omega_{0}}\right)\,i\pi\,\delta(\omega_{0} - {\bf k}\bdot{\bf v}) \;+\; \pi\,e^{2}\,\ov{\Phi}_{1}\,\pd{\ov{\Phi}_{1}^{*}}{t}\,\Delta\,\pd{\delta(\omega_{0} - {\bf k}\bdot{\bf v})}{\omega_{0}} \right]
\end{eqnarray}
so that 
\begin{eqnarray}
   i\,{\bf k}\,\left(\ov{H}_{1}\,\ov{F}_{1}^{*} \;-\frac{}{} \ov{H}_{1}^{*}\,\ov{F}_{1}\right) &=& {\bf k}\,{\bf k}\bdot\pd{F_{0}}{\bf P} \left[ e^{2}\,|\ov{\Phi}_{1}|^{2}\;2\pi\,
  \delta(\omega_{0} - {\bf k}\bdot{\bf v}) + i\,\pi\,e^{2}\left(\ov{\Phi}_{1}^{*}\,\pd{\ov{\Phi}_{1}}{t} - \ov{\Phi}_{1}\,\pd{\ov{\Phi}_{1}^{*}}{t}\right) \pd{}{\omega_{0}}
  \left(\Delta\frac{}{}\delta(\omega_{0} - {\bf k}\bdot{\bf v})\right) \right] \nonumber \\
   &=& \left[ e^{2}\,|\ov{\Phi}_{1}|^{2}\frac{}{} 2\pi\,\delta(\omega_{0} - {\bf k}\bdot{\bf v})\right] {\bf k}\,{\bf k}\bdot\pd{F_{0}}{\bf P} \;\equiv\; \mathbb{D}\bdot\pd{F_{0}}{\bf P},
\end{eqnarray}
where the quasilinear diffusion tensor in oscillation-center phase space is identical to Eq.~\ref{eq:D_QL}, and the partial-time derivative terms
\begin{equation} 
\ov{\Phi}_{1}^{*}\,\pd{\ov{\Phi}_{1}}{t} \;-\; \ov{\Phi}_{1}\,\pd{\ov{\Phi}_{1}^{*}}{t} \;\equiv\; (\gamma - \gamma)\,|\ov{\Phi}_{1}|^{2} \;=\; 0 
\label{eq:phi_phi_dot}
\end{equation}
cancel out under the assumption of a real growth rate $\gamma$. Hence, we conclude from the second-order oscillation-center Vlasov equation becomes
\begin{equation}
    \pd{F_{0}}{\tau} + \pd{\langle F_{2}\rangle}{t} \;=\; \pd{}{\bf P}\bdot\left( \mathbb{D}\bdot\pd{F_{0}}{\bf P}\right) \;\equiv\; \pd{F_{0}}{\tau}
\end{equation}
which implies that the space-averaged second-order oscillation-center Vlasov distribution $\langle F_{2}\rangle$ is independent of time.

We now show that $\langle F_{2}\rangle$ vanishes, by computing the space-averaged of Eq.~(\ref{eq:fF_2}):
\begin{equation}
  \langle f_{2}\rangle \;=\; \langle F_{2}\rangle \;+\; \left\langle \left\{ S_{1},\frac{}{} F_{1}\right\}\right\rangle \;+\; \{ \langle S_{2}\rangle,\; F_{0}\} \;+\; \frac{1}{2} 
  \left\langle\left\{ S_{1},\frac{}{} \{ S_{1},\; F_{0}\} \right\}\right\rangle.
\end{equation}
First, we note that $\langle S_{2}\rangle$ is a function of ${\bf P}$ and, therefore, we find $\{ \langle S_{2}\rangle,\; F_{0}\} \equiv 0$. Next, the term $\left\langle \left\{ S_{1},\frac{}{} F_{1}\right\}\right\rangle$ simultaneously involves the contributions of a resonant term 
$\ov{F}_{1} \sim \Delta$ and a non-resonant term $\ov{S}_{1} \sim (1 - \Delta)$, which implies that $\left\langle \left\{ S_{1},\frac{}{} F_{1}\right\}\right\rangle \equiv 0$ must vanish identically.

Hence, we are left with the relation
\begin{equation}
    \langle F_{2}\rangle \;=\; \langle f_{2}\rangle \;-\; \frac{1}{2} \left\langle\left\{ S_{1},\frac{}{} \{ S_{1},\; f_{0}\} \right\}\right\rangle \;=\; \langle f_{2}\rangle \;-\;
    \frac{1}{2}\,\pd{}{\bf p}\bdot\left(\left\langle \nabla S_{1}\,\nabla S_{1}\right\rangle\bdot\pd{f_{0}}{\bf p}\right),
\end{equation}
which also appears as Eq.~\cite{Dewar:1973}(42) in Dewar's work. First, we find $\left\langle \nabla S_{1}\,\nabla S_{1}\right\rangle \equiv 2\,|\ov{S}_{1}|^{2}\,{\bf k}{\bf k}$, where
\begin{eqnarray}
    |\ov{S}_{1}|^{2} &=& \left|\left( e\,\ov{\Phi}_{1} \;+\; i\,e\,\pd{\ov{\Phi}_{1}}{t}\,\pd{}{\omega_{0}}\right) {\cal P}\left(\frac{1}{(\omega_{0} - {\bf k}\bdot{\bf v})}\right)\right|^{2} \nonumber \\
    &=& e^{2}\,|\ov{\Phi}_{1}|^{2}\;{\cal P}\left(\frac{1}{(\omega_{0} - {\bf k}\bdot{\bf v})^{2}}\right) \;+\; \frac{i}{2}\,e^{2}\left(\ov{\Phi}_{1}^{*}\,\pd{\ov{\Phi}_{1}}{t} - \ov{\Phi}_{1}\,\pd{\ov{\Phi}_{1}^{*}}{t}\right) \pd{}{\omega_{0}}{\cal P}\left(\frac{1}{(\omega_{0} - {\bf k}\bdot{\bf v})^{2}}\right) \nonumber \\
    &=& e^{2}\,|\ov{\Phi}_{1}|^{2}\;{\cal P}\left(\frac{1}{(\omega_{0} - {\bf k}\bdot{\bf v})^{2}}\right) \;\equiv\; -\;e^{2}\,|\ov{\Phi}_{1}|^{2}\;\pd{}{\omega_{0}}{\cal P}\left(\frac{1}{(\omega_{0} - {\bf k}\bdot{\bf v})}\right),
\end{eqnarray}
where we have used Eq.~(\ref{eq:phi_phi_dot}). We therefore find
\begin{equation}
    \frac{1}{2}\,\pd{}{\bf p}\bdot\left(\left\langle \nabla S_{1}\,\nabla S_{1}\right\rangle\bdot\pd{f_{0}}{\bf p}\right) \;=\; \pd{}{\bf p}\bdot\left[ -\;e^{2}\,|\ov{\Phi}_{1}|^{2}\;\pd{}{\omega_{0}}{\cal P}\left(\frac{1}{(\omega_{0} - {\bf k}\bdot{\bf v})}\right)\;{\bf k}{\bf k}\bdot\pd{f_{0}}{\bf p}\right] \;\equiv\; \langle f_{2}\rangle,
    \label{eq:f2_S1S1}
\end{equation}
which is exactly equal to the non-resonant second-order averaged Vlasov distribution $\langle f_{2}\rangle$, defined in Eq.~(\ref{eq:f2_NR}), so that $\langle F_{2}\rangle = 0$.

\subsection{Energy-momentum conservation laws}

The background kinetic energy in oscillation-center phase space is
\begin{equation}
    {\cal E}_{0}(\delta^{2}t) \;=\; \int \frac{|{\bf P}|^{2}}{2m}\; F_{0}({\bf P},\delta^{2}t)\;d^{3}P
\end{equation}
while the second-order oscillation-center total energy is
\begin{eqnarray}
    {\cal E}_{2}(t) &\equiv& \delta^{2} \left[ \frac{|{\bf k}|^{2}}{4\pi}\;|\ov{\Phi}_{1}|^{2} \;+\; \int \left( \frac{|{\bf P}|^{2}}{2m}\;\langle F_{2}\rangle \;-\; \frac{\bf P}{m}\bdot\langle\nabla S_{1}\;F_{1}\rangle \;+\; \frac{1}{2}\pd{}{\bf P}\bdot\left\langle \nabla S_{1}\;\frac{\bf P}{m}\bdot\nabla S_{1}\right\rangle\; F_{0} \right) d^{3}P \right] \nonumber \\
    &=& \delta^{2} \left[ \frac{|{\bf k}|^{2}}{4\pi}\;|\ov{\Phi}_{1}|^{2} \;-\; \int \frac{\bf P}{2m}\bdot\left\langle \nabla S_{1}\;\nabla S_{1}\right\rangle\bdot
    \pd{F_{0}}{\bf P}\;d^{3}P \right],
\end{eqnarray}
where we used $\langle F_{2}\rangle \equiv 0$ and $\langle\nabla S_{1}\;F_{1}\rangle \equiv 0$. By substituting Eq.~(\ref{eq:f2_S1S1}), we see that ${\cal E}_{2}$ is identical to the second-order particle total energy (\ref{eq:E2_def}) and, thus, the energy conservation law is also satisfied in oscillation-center phase space.

Likewise, the background momentum density in osciltation-center phase space is
\begin{equation}
    \vb{\cal P}_{0}(\delta^{2}t) \;=\; \int {\bf P}\;F_{0}({\bf P},\delta^{2}t)\;d^{3}P,
\end{equation}
while the second-order oscillation-center momentum density is
\begin{eqnarray}
    \vb{\cal P}_{2}(t) &\equiv& \delta^{2} \int \left( {\bf P}\;\langle F_{2}\rangle \;-\; \langle\nabla S_{1}\;F_{1}\rangle \;+\; \frac{1}{2}\;\pd{}{\bf P}\bdot\left\langle \nabla S_{1}
    \frac{}{}\nabla S_{1}\right\rangle\; F_{0} \right) d^{3}P \nonumber \\
    &\equiv& -\;\frac{\delta^{2}}{2} \int \left\langle \nabla S_{1}\frac{}{}\nabla S_{1}\right\rangle\bdot\pd{F_{0}}{\bf P}\; d^{3}P,
\end{eqnarray}
where we once again used $\langle F_{2}\rangle \equiv 0$ and $\langle\nabla S_{1}\;F_{1}\rangle \equiv 0$. Once again, by substituting Eq.~(\ref{eq:f2_S1S1}), we see that 
$\vb{\cal P}_{2}$ is identical to the second-order particle momentum density (\ref{eq:P2_def}) and, thus, the momentum conservation law is also satisfied in oscillation-center phase space.

\section{Summary}

By using the Lie-transform perturbation method, we have derived self-consistent quasilinear theories in particle and oscillation-center phase spaces, originally derived by different methods by Kaufman \cite{Kaufman:1972} and Dewar \cite{Dewar:1973}, respectively. Both quasilinear theories preserve the conservation laws of energy and momentum by clearly separating the energy-momentum carried by the resonant particles and the total energy-momentum carried by the waves (which include contributions from the non-resonant particles). 

Future work will expand the recent work by Brizard and Chan \cite{Brizard_Chan:2022}, who presented the Hamiltonian formulation of quasilinear diffusion of resonant particles in a uniform magnetized plasma in the presence of electromagnetic-wave fluctuations, by including the non-resonant particle contributions that preserve the conservation laws of energy and momentum. The generalization to the quasilinear diffusion of relativistic particles in a weakly nonuniform magnetized background plasma that is perturbed by electromagnetic waves of arbitrary frequencies, which was considered in previous papers \cite{Brizard_Chan:2001,Brizard_Chan:2004}, will also be considered.

\acknowledgments
 
The Author dedicates this paper to the memories of Bob Dewar (1944-2024) and Allan Kaufman (1927-2022), who were both pioneers of the Hamiltonian formulation of quasilinear theory and its associated energy-momentum conservation laws, and both were early champions of applications of Lie-transform perturbation methods in plasma physics. This work was partially funded by a grant PHY-2206302 from the National Science Foundation.

\end{document}